\documentclass[onecolumn,showpacs,12pt]{revtex4}
\usepackage{graphicx}
\usepackage{dcolumn}
\usepackage{bm}
\begin{document}
\newcommand{\hs}{\hspace*{0.5cm}}
\newcommand{\vs}{\vspace*{0.5cm}}
\newcommand{\be}{\begin{equation}}
\newcommand{\ee}{\end{equation}}
\newcommand{\bea}{\begin{eqnarray}}
\newcommand{\eea}{\end{eqnarray}}
\newcommand{\ben}{\begin{enumerate}}
\newcommand{\een}{\end{enumerate}}
\newcommand{\bde}{\begin{widetext}}
\newcommand{\ede}{\end{widetext}}
\newcommand{\nn}{\nonumber}
\newcommand{\crn}{\nonumber \\}
\newcommand{\Tr}{\mathrm{Tr}}
\newcommand{\noi}{\noindent}
\newcommand{\al}{\alpha}
\newcommand{\la}{\lambda}
\newcommand{\bet}{\beta}
\newcommand{\ga}{\gamma}
\newcommand{\va}{\varphi}
\newcommand{\om}{\omega}
\newcommand{\pa}{\partial}
\newcommand{\+}{\dagger}
\newcommand{\fr}{\frac}
\newcommand{\bc}{\begin{center}}
\newcommand{\ec}{\end{center}}
\newcommand{\Ga}{\Gamma}
\newcommand{\de}{\delta}
\newcommand{\De}{\Delta}
\newcommand{\ep}{\epsilon}
\newcommand{\varep}{\varepsilon}
\newcommand{\ka}{\kappa}
\newcommand{\La}{\Lambda}
\newcommand{\si}{\sigma}
\newcommand{\Si}{\Sigma}
\newcommand{\ta}{\tau}
\newcommand{\up}{\upsilon}
\newcommand{\Up}{\Upsilon}
\newcommand{\ze}{\zeta}
\newcommand{\ps}{\psi}
\newcommand{\Ps}{\Psi}
\newcommand{\ph}{\phi}
\newcommand{\vph}{\varphi}
\newcommand{\Ph}{\Phi}
\newcommand{\Om}{\Omega}

\title{Early Universe  in   the  $ \mathrm{SU}(3)_L \otimes \mathrm{U}(1)_X$ electroweak models
}

\author{Hoang  Ngoc  Long}
\email{hnlong@iop.vast.vn} \affiliation{Institute of Physics,
Vietnam Academy of Science and Technology, \\
10 Dao Tan, Ba Dinh, Hanoi, Vietnam}

\date{\today}

\begin{abstract}
We present status of the 3-3-1 models and their implications to  cosmological evolution such as inflation, phase transitions and sphalerons.   The models can deal not only with the  issues such as   neutrino physics,  dark matter, etc,  but  they are   also  able  to  provide quite good
 agreement with the Standard Cosmology:   the
inflation happens at the GUT scale, while  phase transition has two sequences corresponding two steps
of symmetry breaking in the models,  namely: $SU(3) \rightarrow SU(2)$ and $SU(2) \rightarrow U(1)$.   Some
bounds on the model parameters are obtained: in the RM331,  the  mass of the heavy neutral Higgs boson is fixed  in the range:
$285.56\, {\rm GeV}<M_{h_2}< 1.746 \, {\rm TeV}$, and for  the  doubly charged scalar:
 $3.32\,  {\rm TeV} <M_{h_{--}}< 5.61\,  {\rm TeV}$.

\end{abstract}

\maketitle

\section{\label{intro} Introduction}
\label{intro}
It is well known that our Universe content is 68.3\% of Dark Energy (DE), 26.8\% of Dark Matter (DM) and of
 4.9\% of luminous matter \cite{plank}.  With the unique fact of  accelerating  Universe,   the core origin of
   Dark Energy is still under question, while the existence of Dark Matter is unambiguous. According to the Standard
   Cosmology,  in the moment at $10^{-36} s$ after the Big Bang (BB), there was  inflation, and our Universe has been expanded
   exponentially. The inflationary scenario solves a number of problems such as the Universe's flatness,
 horizon, primordial monopole, etc. It is well known that there is no anti-matter in our Universe,
  or other word speaking: at present there exists a Baryon Asymmetry of Universe (BAU).
The baryon number vanishes ($n_B=0$) at the BB, and this conflicts with the present  BAU. Nowadays,
    the BAU is one of the greatest challenges in Physics and any physical model has to give  an explanation.
      The BAU is realized  if three   Sakharov's conditions are satisfied  \cite{sakharov,mkn}
   \ben
   \item $B$ violation,
   \item $C$ and $CP$ violations,
   \item deviation from thermal equilibrium.
   \een

Over the half of  Century, the  Standard Model (SM)  of the electromagnetic, weak and strong interactions
successfully possesses a great experimental examinations
and stands for future development. Despite its great success, the model still contains a number of unresolved
problems such as the generation number of quarks and leptons, the neutrino mass and mixing,  the electric charge
quantization, the existence of about one quarter of DM,  etc.
The aforementioned problems require that the SM must be extended.

Among the extensions beyond the SM, the models based on $\mathrm{SU}(3)_C\otimes \mathrm{SU}(3)_L
\otimes \mathrm{U}(1)_X$ (3-3-1) gauge group  \cite{ppf, flt} have some interesting features including
 the ability to explain the generation problem \cite{ppf, flt} and the electric charge quantization \cite{chargequan}.
 It is noted that in this scheme the gauge couplings can be unified
at the scale of order TeV\emph{ without  supersymmetry} \cite{uni}.
 The 3-3-1 models have two interesting properties needed for the mentioned aim, namely: first, the lepton-number violation due to the fact that lepton and anti-lepton are put in the triplet \cite{changlong}. Second, one generation of quarks  transforms differently from other two. This leads to the flavor changing neutral current at the tree level mediated by new $Z'$ gauge boson \cite{longvan}.

The 3-3-1 models have been considered in aspects of collider physics \cite{331LHC,331ep,Higgsdecay331,cataenoEPJC},
 muon anomalous magnetic moments \cite{g2muon}, neutrino physics
\cite{neutrino331},  DM \cite{s331,dm331}.... In this review I will concentrate on Early Universe aspects of the models.

This paper is organized as follows. In Sec. \ref{model} we give a brief  review of the 3-3-1 models and their modified versions.
 In Sec. \ref{cosmology}, the cosmological inflation in the supersymmetric economical 3-3-1 model is presented.
  In Sec. \ref{ept}, we investigate the structure of the electroweak phase transition (EWPT) sequence
  in the  3-3-1 models with minimal Higgs sector, namely the reduced minimal 3-3-1 model (RM331) and the economical 3-3-1 model   (E331), find the parameter ranges where the EWPTs are the strongly first-order to provide B violation necessary for
  baryogenesis, and show the constraints on the mass of the charged Higgs boson.
   Section \ref{spha} is devoted for sphalerons in the reduced minimal
    3-3-1 model.
  Finally,  in Sec. \ref{conclus} we give conclusion on the possibility
  to describe cosmological evolution in the framework of the 3-3-1 models.

\section{The models\label{model}}
In  the mentioned models, the strong interaction keeps  the same as in the SM, while
the electroweak part associated with $  \mathrm{SU}(3)_L \otimes \mathrm{U}(1)_X$
has two diagonal generators $T_3$ and $T_8$ from which the electric charge operator is based on
\be Q = T_3 + \bet T_8 + X. \label{qoperator}
\ee
The coefficient (=1) at the $T_3$ is defined to make the 3-3-1 models  embed the SM.
 The lepton arrangement will define the parameter $\bet$ which distinguishes  two main versions:
the minimal version with $\bet = \sqrt{3}$ and the version with neutral leptons/neutrinos
$\bet = - 1/\sqrt{3}$ at the bottom of the triplet.

\subsection{The minimal 3-3-1 model\label{mmodel}}
The minimal version \cite{ppf} contains lepton triplet in the form
\be f_L = (\nu_l,\,  l,\,  l^c)_L^T \sim  (1, 3, 0).
 \label{leptonm}
\ee
Two first quark generations are in anti-triplet and the third one is in
titriplet:
  \be Q_{iL} = \left(d_{iL},-  u_{iL} ,  D_{iL}
 \right)^T \sim \left(3, \bar{3} , -\fr{1}{3}\right), \label{q} \ee
\[ u_{iR}\sim (3, 1, 2/3), d_{iR}\sim (3, 1, -1/3),
D_{iR}\sim (3, 1, -4/3),\ i=1,2,\] \[ Q_{3L} = \left(
  u_{3L},  d_{3L},  T_{L} \right)^T \sim (3, 3, 2/3),\]
\[ u_{3R}\sim (3, 1, 2/3), d_{3R}\sim (3, 1, -1/3), T_{R}
\sim (3, 1, 5/3).\]
To provide  masses for all quarks and lepton, the Higgs sector needs three scalar triplets and one sextet:
 \bea \chi & = &\left(  \chi_1^{-},  \chi_2^{--},  \chi_3^{0}
  \right)^T \sim (1, 3, -1),\label{mh1}\\
 \eta & =& \left(
   \eta_1^0, \, \eta_2^-, \,  \eta^+_3  \right)^T \sim (1, 3, 0),\crn
 \rho & =& \left( \rho_1^{+},  \rho_2^{0},  \rho_3^{++}
  \right)^T \sim (1, 3, 1),\crn
S & \sim &
 (1, 6, 0).\nn\eea
  with VEV: $\langle  \rho_2^0 \rangle = v/ \sqrt{2}, \langle \eta_1^0
\rangle = u/ \sqrt{2}$ ,  $\langle \chi_3^{0} \rangle =  \om/
\sqrt{2} $ and $   \langle S^0_{23} \rangle =v' /\sqrt{2}$.

The gauge sector of this model contains five new gauge bosons: one neutral $Z'$ and two bileptons carrying lepton number 2:
$Y^\pm$ and $X^{\pm \pm}$.
In (\ref{leptonm}), lepton and antilepton lie in the same triplet, and this leads to lepton number
violations in the model. Hence, it is better to deal with a new conserved charge $\mathcal{L}$ commuting with the gauge symmetry \cite{changlong}
\be  L = \fr{4}{\sqrt{3}} T_8 + \mathcal{L}.
 \label{leptonct}
\ee
The exotic quarks $T$ and $D_i$ have the electric charges, respectively, $5/3$ and $-4/3$
and  carry both baryon and lepton numbers $ L = \pm 2$.

 The singly charged bilepton is responsible for the wrong muon decay
\[ \mu \rightarrow e + \nu_e + \tilde{\nu_\mu},\]
while the doubly charged bilepton with decay
\[ X^{--}  \rightarrow l\,  l\]
provides four leptons at the final states which is characteristic feature of the model.
The model provides an interesting prediction for the Weinberg angle
\[ \sin^2 \theta_W (M_{Z'}) \leq \fr 1 4.\]
Besides the complication in the Higgs sector, the model also has one problem that it
losses perturbative property at the scale above  5 TeV \cite{dias}.

The above Higgs sector is complicated; and recently it is reduced to the minimal with only
two Higgs triplets \cite{rm331,s331}. If the triplet $\rho$ and $\chi$ are used then the model is called
reduced minimal 3-3-1 model \cite{rm331}, while $\rho$ is replaced by $\eta$ then it is
called simple 3-3-1 model (S331) \cite{s331}.

It has been recently shown that due to the $\rho$ parameter and the Landau pole, the minimal and its
reduced version should be ruled out \cite{dsi}. It is noted that the RM331 has nonrenormalizable
 effective interactions, so situation has to be considered carefully.

\subsection{The  3-3-1 model with right-handed neutrinos\label{rmodel}}
  Leptons are in triplet ~\cite{flt}: \be f^{a}_L = \left(
  \nu^a_L,  e^a_L,  (N_L)^a
  \right)^T \sim (1, 3, -1/3), e^a_R\sim (1, 1, -1), \label{l21} \ee
where  $ a = 1, 2, 3$ is a generation index and $N_L$ can be right-handed neutrino or neutral lepton.
 Two first generations
of quarks are in antitriplets, and the third one is in triplet:
\be Q_{iL} = \left(  d_{iL}, -u_{iL},  D_{iL}
 \right)^T \sim (3, \bar{3}, 0), \label{q} \ee
\[ u_{iR}\sim (3, 1, 2/3), d_{iR}\sim (3, 1, -1/3),
D_{iR}\sim (3, 1, -1/3),\ i=1,2,\] \[
 Q_{3L} = \left(  u_{3L},  d_{3L},  T_{L}
 \right)^T \sim (3, 3, 1/3),\]
\[ u_{3R}\sim (3, 1, 2/3), d_{3R}\sim (3, 1, -1/3), T_{R}
\sim (3, 1, 2/3).\]
The model with  neutral lepton/neutrino ($\bet = - 1/\sqrt{3}$) needs three scalar triplets to
provide all fermions masses and the same for
  spontaneous symmetry breaking (SSB):
 \bea \chi & = &\left( \chi^0, \chi^-,  \chi^{,0}
 \right)^T \sim (1, 3, -1),\crn
 \rho & =& \left( \rho^+,  \rho^0,  \rho^{,+}
 \right)^T \sim (1, 3, 2),\label{h1}\\
\eta & =& \left(  \eta^0, \eta^-,  \eta^{,0}
 \right)^T \sim (1, 3, -1).\nn \eea

 The exotic quarks $T$ and $D_i$ have  electric charges as usual one, i.e.,  $2/3$ and $-1/3$, respectively, and
 carry both baryon and lepton numbers $ L = \pm 2$. The new gauge bosons are: the  neutral $Z'$ and two bileptons carrying lepton number 2:
$Y^\pm$ and $X^{0}$.  The neutral bilepton $X^{0}$ is non-Hermitian and is responsible for neutrino oscillation
\cite{longinami}.

Note that two Higgs triplets $\eta$ and $\chi$ have the same structure, so ones can reduce number of Higgs triplets from three to two, namely
 we can use only $\rho$ and $\chi$ to produce masses for quarks and leptons; and resulting model is called economical 3-3-1 model \cite{ecn331}.
 As in the RM331, the nonrenormalizable interactions, in this case, are needed for production of
quark masses \cite{ecn331}.

\section{\label{cosmology}Cosmological inflation in the supersymmetric economical
 3-3-1 models}

The discovery of the 2.7$K$
microwave background radiation arriving from the farthest reaches
of the Universe,  gained  widespread acceptance, is positive point of   the hot-universe theory,
where the inflationary scenario \cite{infsce,lind383} plays very important role.
Cosmological inflation (CI) can give solutions for above mentioned problems, hence
it  is a possible theory of the origin of all structures in the Universe, including ourselves!

With above reasons,  any beyond standard model has to have the cosmological
inflation happened at the  interval of  $10^{-36} - 10^{-34} s$ after the BB. With that moment,
the energy scale of CI is about $10^{15} $ GeV. In \cite{cos1}, the CI was considered
in the framework of the supersymmetric economical 3-3-1 model (SE331), and a reason is the
following: the E331 is very simple, but there is no candidate for inflaton - a key element of CI.
The SE331 has some advantages such as there are more scalar fields
which can play a role of the inflaton, and  the Higgs sector is very constrained.

A supersymmetric version of the minimal 3-3-1 model has been constructed in ~\cite{msusy} and its scalar
sector was  studied in ~\cite{duongma}. Lepton masses in the framework
of the above-mentioned model were presented in
\cite{leptonmassm331}, while potential discovery of supersymmetric
particles  was studied in \cite{consm331}. In \cite{longpal}, the
$R$-parity violating interaction was applied for  instability of
the proton. A supersymmetric RM331 was presented in \cite{srm331}.

The supersymmetric version of the 3-3-1 model with right-handed
neutrinos has already been constructed in~\cite{s331r}. The scalar
sector was considered in~\cite{scalarrhn} and neutrino mass was
studied in ~\cite{marcos}.
A supersymmetric version of the economical 3-3-1 model has been
constructed in  \cite{susyeco}. Some interesting features such as
Higgs bosons with masses equal to that of the gauge bosons: the
$W$ ($m^2_{\varrho_{1}^+}=m^2_W$) and the bileptons $X$ and $Y$
($m^2_{\zeta^\pm_4} = m^2_Y$), have been pointed out in
\cite{higph}. Sfermions in this model have been considered in
\cite{jhep}. In \cite{jhep2} it was shown that bino-like
neutralino can be a candidate for DM.

In \cite{cos1}, the authors have
constructed a hybrid inflationary scheme based on a realistic
supersymmetric $ SU(3)_C \otimes SU(3)_L \otimes U(1)_X $
model by adding a singlet superfield $\Phi$ which plays the role
of the inflaton, namely the inflaton superfield.

 We remind that  the existence of a $U(1)_Z$ does not
belong to the MSSM  and it   spontaneously  breaks  down  at the
scale $M_X$ by  Higgs  superfield   $\phi$, which is singlet under the MSSM.
The inflaton superfield couples with this pair of
Higgs superfields. Therefore, the additional global supersymmetric
renorrmalizable superpotential for the inflation sector is chosen
to be~\cite{infpot,GDvali} \be
W_{inf}(\Phi,\chi,\chi^\prime)=\alpha \Phi \chi \chi^\prime -\mu^2
\Phi.\label{Poten}\ee  The superpotential given by (\ref{Poten})
is the most general potential consistent with a continuous $R$
symmetry under which $\phi \rightarrow e^{i \ga} \phi,  \ W
\rightarrow e^{i \ga} W $, while the product $\chi \chi^\prime$ is
invariant ~\cite{GDvali,linrio}.

By a suitable redefinition of complex
 fields  $\mu^2,\alpha$ are chosen to be
 positive real constants , and the ratio $\frac{\mu} {\sqrt{ \alpha}}$ sets the $U(1)_Z$
 symmetry breaking scale $M_X$. The most general
 superpotential consistent with a continuous $R$-symmetry
  is given by
\be W_{tot}=W_R + W_{inf}(\Phi,\chi,\chi^\prime).
\label{Poten1}\ee
With the superpotential
given in (\ref{Poten}),  the Higgs scalar potential takes the form
 \[ V_{tot} =
\Sigma_{i}|F_i|^2 + \frac{1}{2}\sum_{\alpha}|D_\alpha|^2+
V_{soft},\]  where $i$ runs from 1 to the total number of the
chiral superfields in $W_{tot}$, while $V_{soft}$ contains all the
soft terms generated by supersymmetry breaking at the low energy.

Hence, the Higgs potential becomes \bea
V_{tot}&=&|\mu_\chi  +\alpha \Phi |^2|\chi^\prime|^2+ |\mu_\chi
+\alpha \Phi |^2|\chi|^2+|\alpha \chi
\chi^\prime -\mu^2|^2+\crn & & + |\mu_\rho \rho|^2+|\mu_{\rho^\prime
}\rho^\prime|^2 + \frac{1}{2}\sum_{\alpha}|D_\alpha|^2+
V_{soft}.\nn\eea
 The first derivatives $\frac{\pa V_{tot}}{\pa \rho}, \frac{\pa
V_{tot}}{\pa \rho^\prime}$ are independent of $\chi, \chi^\prime,
\Phi$, and  the fields $\rho, \rho^\prime$ will stay
in their minimum independently of what the fields $\chi,
\chi^\prime, \Phi$ do. If  we are mainly interested
in what is happening above the electroweak scale, and hence we do
not take into account the dimensional Higgs  multiplets $\rho,
\rho^\prime$. Then, the Higgs scalar potential is  given by
\bea V_{inf}&=&|\mu_\chi +\alpha \Phi |^2|\chi^\prime|^2+
|\mu_\chi +\alpha \Phi |^2|\chi|^2+|\alpha \chi \chi^\prime
-\mu^2|^2+ \crn & & + \frac{1}{2} \left( g\sum_a\chi^* T^a
\chi\right)^2+\frac{1}{2} \left(g\sum_a\chi^{\prime*} T^a
\chi^\prime\right)^2. \label{Poten2} \eea
  Let us denote
 \be \mu_\chi
+\alpha \Phi \equiv \beta S,\label{nha}\ee where $\bet$ is some
constant and $S$ is a new field,   the Higgs potential
(\ref{Poten2}) can be rewritten as \bea V_{inf} &=& \beta^2
|S|^2\left( |\chi|^2+ |\chi^\prime|^2\right)+|\alpha \chi
\chi^\prime -\mu^2|^2 \crn && + \frac{1}{2} \left( g\sum_a\chi^*
T^a \chi\right)^2+\frac{1}{2} \left(g\sum_a\chi^{\prime*} T^a
\chi^\prime\right)^2. \nn\eea

 When $D$ term vanishes along its  direction, the potential
 contains only $F$ term and has the form
 \be
V_{inf}=\beta^2 |S|^2\left( |\chi|^2+
|\chi^\prime|^2\right)+|\alpha \chi
\chi^\prime-\mu^2|^2.\label{th271}
 \ee
From (\ref{th271}), it is clear that  $V_{inf}$ has an unique
supersymmetric minimum corresponding to
 \bea
 \langle  S  \rangle & = & 0, \crn
 M_X & \equiv &  \langle  \chi   \rangle  =  \langle
 \chi^\prime \rangle=\frac{\mu}{\sqrt{\alpha}}.
\label{th272} \eea  The ratio $\frac{\mu}{\sqrt{\alpha}}$ sets the
$U(1)_Z$ symmetry breaking $M_X$, but  Eq. (\ref{th272}) is global
minimum, and supersymmetry is not violated~\cite{GDvali}. Hence,
inflation can take  place but supersymmetry is not broken. This is $F$ term inflation ~\cite{Jen}.

 We assume  that the initial
 value for the inflaton field is much greater than its critical
 value $S_c$. For $|S|> |S_c|\equiv \frac{\mu}{\sqrt{\alpha}} $  the potential is
 very flat in the $|S|$ direction, and the $\chi, \chi^\prime $
 fields settle down to the local minimum of the potential,
 $\chi=\chi^\prime=0$, but it does not drive $S$ to its minimum
 value.  The universe is dominated by a nonzero vacuum energy density,
 $V_0^{\frac{1}{4}}=\mu$, which can lead to an exponential expanding,  inflation starts,
  and supersymmetry is broken.

   By the Coleman-Weinberg formula in \cite{Coliman}, at the one-loop level, the effective potential along the
 inflaton direction is given by
 \[
 \Delta V=\frac{1}{64\pi^2}\sum_i ( -1)^F m_i^4
 \ln\left( \frac{m_i^2}{\Lambda^2}\right),
 \]
where  $F= -1$  for the fermionic fields and
 $F=1$ for the bosonic fields. The coefficient $(-1)^F$ shows that bosons and fermions
 give opposite contributions.  The sum runs over each degree of
 freedom $i$ with mass $m_i$ and $\Lambda$ is a renormalization scale.

The effective potential (along the inflationary trajectory $S
> S_c,\chi=\chi^\prime=0$) is given by
 \bea
  V_{eff} (S)&=& \mu^4+\frac{3}{16\pi^2}\left[ 2 \beta^4 \frac{\mu^4}{\alpha^2}
  \ln\frac{\beta^2 |S|^2}{\Lambda^2}
 +\left( \beta^2 |S|^2+ \alpha \mu^2\right)^2
  \ln \left( 1+\frac{\alpha\mu^2}{\beta^2 |S|^2}\right)+ \right.
  \crn && \left.
   + \left( \beta ^2|S|^2- \alpha \mu^2\right)^2 \ln \left( 1-\frac{\alpha\mu^2}{\beta^2
   |S|^2}\right)\right]\label{Veff}
  \eea

 It is to be noted that  for  $S > S_c $, the universe is dominated by the false
 vacuum  energy $\mu^4$. When $S$ field drops to $S_c$, then the
 GUT phase transition happens. At the end of inflation, the inflaton field does not
 need to coincide  with the GUT phase transition. The end of inflation
 can be supposed to be on a region of the potential which
 satisfies the flatness conditions (see, for example, ~\cite{lythrioto})
 \be
 \epsilon  \ll1 , \,  \eta   \ll 1,
 \label{cond1}\ee
 where  we have used the conventional notations
 \be
 \epsilon  \equiv \frac{M_P^2}{16 \pi}\left(\frac{V^\prime}{V}
 \right)^2,\ ,  \eta \equiv \frac{M_P^2}{8\pi} \frac{V^{\prime \prime}}{V }
\label{cond11},\ee
 where  primes denote a derivative with respective  to $S$.

To compare with observational  COBE data, we use   the slow-roll approximation with parameters:
 $\ep$ and $\eta$.
 The first condition in (\ref{cond1}): $\epsilon \ll1$ indicates that the density $\rho$
 is close to $V$ and is slowly varying. As a result,  the  Hubble parameter  $H$ is
 slowly varying, which implies that one can write $a\propto
 e^{Ht}$ at least over a Hubble time or so. The second condition $\eta \ll 1$
 is a result of the first condition plus the slow-roll
 approximation. The  conditional   phase may end before the GUT
 transition if the flatness conditions (\ref{cond1})  are violated at some
  point  $S > S_c$.

Let us denote a dimensionless
variable \be y \equiv  \frac{\beta |S|}{\alpha S_c
}.\label{th721}\ee
  Imposing  the condition $\alpha =\beta$,  which means that $|\Phi| \approx |S| \gg \mu_\chi$,
    we get then
   \bea
 \epsilon &=&\left(\frac{3\alpha^2 M_P}{4\pi^2 M_X}\right)^2\frac{1}{16
 \pi}  \left[
  y\left( y^2-1\right)\ln \left( 1-\frac{1}{y^2}\right)+
 \right.\crn && \left.
 + y\left( y^2+1\right)\ln \left( 1+\frac{1}{y^2}\right)\right]^2, \crn
  \eta &=&\left(\frac{\alpha M_P}{4\pi M_X}\right)^2\frac{3}{2\pi }
  \left[(3y^2+1)\ln\left(1+\frac{1}{y^2} \right)+(3y^2-1)\ln\left(1-\frac{1}{y^2} \right)
   \right] \label{cond3}\eea

The chaotic inflation driven by the $\phi^3$ is in good agreement with the WMAP data, while
for the $\phi^4$ potential, the situation is negative.

The above model cannot resolve the horizon/flatness problems of the BB cosmology and violates
the slow-roll conditions $\eta \ll 1$ (the $\eta$  problem). To deal with these problems, we should
consider the $F$-term inflation with minimal K\"{a}hler potential.

 The $F$-term inflation with  K\"{a}hler potential is defined by
 \be
W_{stand}(\Phi,\chi,\chi^\prime)= \al \widehat{S}
\left(\widehat{\chi} \widehat{\chi^\prime} -M_X^2
\right).\label{sugra2}\ee
 Keeping in mind  that
  $K = \sum_\al \left|\phi_\al \right|^2$, we obtain the scalar potential
 \bea V^m_F & = & 2 \al^2 S^2 \phi^2 \left [1 + \fr{S^2 +
 2\phi^2}{m_P^2} + \fr{(S^2 +
 2\phi^2)^2}{2 m_P^4} \right]+\crn
&&+\al^2(\phi^2 -M_X^2)^2\left( 1 + 2\fr{\phi^2}{m_P^2} +
\fr{S^4}{2 m_P^4} + 2\fr{\phi^4}{m_P^4}\right) + \cdot \cdot
\cdot,\label{sugra3} \eea where we have assumed that $|\phi|^2 =
|\phi^\prime|^2$ .

 Let us consider how  does this factor change the result. As we know, the
slow-roll parameter  is defined as \[ \eta = m_p^2 \left(
\frac{V^{\prime \prime}}{V}\right),  \]
where the prime refers to derivative with respect to $S$.  The supergravity scalar potential for $S>S_c$
is given by \be V_o=\al^2 M_X^4 + \frac{\al^2 M_X^4}{2m_p^4}S^4.
\label{sugra4} \ee  From (\ref{sugra4}), it follows
derivative of $V$:  $ V^{\prime \prime }\simeq \frac{\al^2
M_X^4}{2 m_p^4} S^2$, and   $\eta = \frac{1}{2 m_p^2} S^2
\ll 1$. Therefore,  the $\eta$-problem is overcome.

The potential given in (\ref{sugra4}) does not contain a term which
can drive $S$ to its minimum value, so we have to
consider the effective potential.
In this case,  the spectral index $n$  is given by
 \bea
n&=& 1-6\epsilon + 2 \eta\crn
&=& 1-\frac{3\al^2}{512 \pi^3 \zeta x^4}
\left[x^2(16+9\al^2)-54\al^2 \zeta +6x^8(-40+9 \al^2)\zeta
+16x^4(-5+9\al^2)\zeta \right],  \label{spesugra} \eea
where $\zeta\equiv \frac{M_X^2}{M_P^2}$.

Taking into account  the
WMAP data,  we conclude that the value of e-folding number $N_Q$ must be larger than 45, and  get
 bounds on  the values of  coupling $\al$ and   $\zeta$, which are presented
 in Table \ref{thf}:

\begin{table}[h]
\caption{
  Bounds on the parameter $\zeta$ and coupling $\al$ followed by the  WMAP data .}
\bc
\begin{tabular}{|c|c|c|c|c|}
  \hline
  $\al$ & $10^{-3}$ & $ 10^{-4}$  &  $10^{-5}$  &  $10^{-6}$ \\
  \hline
  $\zeta$ &  $25\times 10^{-6}$ &  $25 \times 10^{-7}$ &  $ 25 \times 10^{-9}$
  &   $3 \times 10^{-11}$\\
  \hline
\end{tabular}\label{thf}
\ec
\end{table}

It is interesting to note that due the   inflaton
with mass in the GUT scale, the model can provide masses for neutrino different
from ones without  inflationary scenario.  With the help of the lepton-number-violating
interactions among the inflaton and right-handed neutrinos, the  non-thermal
leptogenesis scenario is followed \cite{huong2}.

In recent work \cite{hdkt}, the authors have considered the inflationary scenario and leptogenesis in newly
proposed 3-3-1-1 model. Here, the scalar field that spontaneously breaks the $U(1)_N$ symmetry plays
a role of  inflaton.

To finish this section, we emphasize that the 3-3-1 models can provide the inflationary scenario or  cosmological evolution of  our Universe.

\section{\label{ept}Electroweak phase transition in 3-3-1 models}

It is known that if baryon number is conserved and is equal to zero, it will equal to zero forever.
If baryon number does not satisfy any conservation law, it vanishes in the state
of thermal equilibrium. Therefore we need the third Sakharov's  condition. The second condition
 is  appropriate  for ensuring a different decay rate for particles and antiparticles \cite{mkn}.
 The electroweak phase transition is the transition
between symmetric phase to asymmetric  phase in order to generate mass for particles.
Hence, the phase transition is related to the mass of the Higgs boson \cite{mkn}.

In the basic model of particles, the first and second conditions can be satisfied,
but conditions on thermal imbalance is difficult to satisfy. So the analysis of
the third condition is the only approach at present in order to explain the baryon asymmetry.

Why is the first order phase transition?
For very large temperature, the effective potential has only one minimum at the zero.
As temperature drops below the critical temperature ($T_c$), the
second minimum appears. If the two minimums are separated by a
potential barrier, the phase transition occurs with bubble
nucleation. Inside the bubbles,  the scalar field acquires a
nonzero expectation value. If the bubble nucleation rate
exceeds the universe's  expansion rate,
 the bubbles collide and eventually fill all space. Such a transition is called
 the first order phase transition. It is very violent and one can expect large deviations
  from thermal equilibrium \cite{mkn}. The other possible scenario takes place if
the two minimums are never separated by a potential barrier. The
phase transition is a smooth transition or the second order phase
transition.

\subsection{Phase transition in reduced minimal 3-3-1 model}

 For the SM, although the EWPT strength is larger than unity at the electroweak scale,
  it is still too weak for the mass of the Higgs boson to be compatible with current
  experimental limits \cite{mkn, SME}; this suggests that electroweak baryogenesis (EWBG) requires new physics beyond the SM at the weak  scale \cite{BSM}. Many extensions such as the two-Higgs-doublet model or Minimal Supersymmetric Standard Model  have a more strongly first-order phase transition and the new sources of CP violation, which are necessary to account
     for the BAU; triggers for the first-order phase transition in these models are heavy bosons or DM
      candidates \cite{majorana, thdm, ESMCO, elptdm}.

To start, let us consider the hight-temperature effective potential
\[
V_{eff}=D.(T^2-{T'}^2_0){v}^2-E. Tv^3+\frac{\lambda_T}{4}v^4,
\]
where $v$ is the VEV of Higgs. In order to have the strongly first-order phase transition,
 the strength of phase transition has to be  larger than 1, i.e., $\frac{v_c}{T_c}\ge 1$.

The phase transition has been firstly  investigated   in the SM. But the difficulty of the SM is that the strength
of the first-order electroweak phase transition, which must be larger than 1 at the
 electroweak scale, appears too weak for the experimentally allowed mass of
 the SM scalar Higgs boson \cite{mkn,SME}. Therefore, it seems that EWBG
  requires a new physics beyond the SM at weak scale \cite{BSM}.

With the discovery of the Higgs boson, the study of phase transitions in the particle
 models is simplified: only to determine the order of phase transition.
This opens a lot of hope for the extended models in examining the electroweak
 phase transition.

The 3-3-1 models  must have at least two Higgs
triplets \cite{ecn331,rm331}. Therefore, the number of bosons in
the 3-3-1 models will many more than in the SM and symmetry
breaking structure is different to the SM.

The physical scalar spectrum of the RM331 model is composed by a doubly charged
 scalar $h^{++}$ and two neutral scalars $h_1$ and $h_2$ \cite{rm331}.
 These new particles and exotic quarks can be triggers for the first order phase transition.

From the Higgs potential we can obtain $V_{0}$ that depends on
VEVs as the following \[ V_{0}(v_{\chi},
v_{\rho})=\mu^2_1v_{\chi}^2+\mu^2_2v_{\rho}^2
+\lambda_1v_{\chi}^4+\lambda_2v_{\rho}^4+(\lambda_3
+\lambda_4)v_{\chi}^2v_{\rho}^2.\]

The effective potential being a function of VEVs and
temperature has the form \[
V=V_{0}(v_{\chi}, v_{\rho})+\sum M^2_{boson}(v_{\chi}, v_{\rho})
W^{\mu}W_{\mu}+\sum m_{fermion}(v_{\chi},
v_{\rho})\overline{f^c_{L}}f^c_{L}. \]

Averaging over space, we obtain
\[ V=V_{0}(v_{\chi}, v_{\rho})+\sum M^2_{boson}(v_{\chi},
v_{\rho}) \left\langle W^{\mu}W_{\mu}\right\rangle+\sum
m_{fermion}(v_{\chi}, v_{\rho}) \left\langle
\overline{f^c_{L}}f^c_{L}\right\rangle \] where $W^\mu$ runs over
all gauge fields. The RM331 has the following  gauge bosons: Two like the SM
bosons $Z_1$, $W^{\pm}$
 and the new  heavy neutral boson $Z_2$, the singly and doubly charged
  boson $U^{\pm\pm}$ and $V^{\pm}$.
 Two doubly charged Higgs $h^{++}$ and  $h^{--}$, one heavy neutral Higgs $h_2$ and
 one like-SM Higgs $h_1$. Using Bose-Einstein and Fermi-Dirac distributions
 for bosons and fermions, we can obtain the effective potential in the RM331 as follows
\bea V_{eff}^{RM331}&= &V_{0}(v_{\chi},
v_{\rho})+\frac{3}{64\pi^2}
\left(m^4_{Z_1}\ln\frac{m^2_{Z_1}}{Q^2}+m^4_{Z_2}\ln\frac{m^2_{Z_2}}{Q^2}
+2m^4_W\ln\frac{m^2_W}{Q^2}\right. \crn &-& \left. 4m^4_t
\ln\frac{m^2_t}{Q^2}-12m^4_Q\ln\frac{m^2_Q}{Q^2}\right)\crn
&+&\frac{1}{64\pi^2}\left(m^4_{h_2}\ln\frac{m^2_{h_2}}{Q^2}+2m^4_{h^{++}}
\ln\frac{m^2_{h^{++}}}{Q^2}\right)\crn
&+&\frac{3}{64\pi^2}\left(2m^4_U\ln\frac{m^2_U}{Q^2}+2m^4_V
\ln\frac{m^2_V}{Q^2}\right)\crn
&+&\frac{T^4}{4\pi^2}\left[F_{-}\left(\frac{m_{h_2}}{T}\right)+2F_{-}\left(\frac{m_{h^{++}}}{T}\right)\right]\crn
&+&\frac{3T^4}{4\pi^2}\left[4F_{+}\left(\frac{m_t}{T}\right)+12F_{+}\left(\frac{m_Q}{T}\right)\right]\crn
&+&\frac{3T^4}{4\pi^2}\left[F_{-}\left(\frac{m_{Z_1}}{T}\right)+F_{-}\left(\frac{m_{Z_2}}{T}\right)
+2F_{-}\left(\frac{m_W}{T}\right)\right.\crn &+& \left.
2F_{-}\left(\frac{m_U}{T}\right)
+2F_{-}\left(\frac{m_V}{T}\right)\right], \nn\eea where \bea
F_{\mp}\left(\frac{m_{\phi}}{T}\right)&=&\int^{\frac{m_{\phi}}{T}}_0\alpha
J^{(1)}_{\mp}(\alpha,0)d\alpha \crn
J^{(1)}_{\mp}(\alpha,0)&=&2\int^{\infty}_{\alpha}\frac{(x^2-\alpha^2)^{1/2}}{e^{x}\mp
1}dx. \nn\eea

The effective potential can be rewritten as follows
\[ V_{eff}=V_0+V^{hard}_{eff}+V^{light}_{eff},\]
where \bea
V^{hard}_{eff}&=&\frac{3}{64\pi^2}\left(m^4_{Z_2}\ln\frac{m^2_{Z_2}}{Q^2}
+m^4_{h_2}\ln\frac{m^2_{h_2}}{Q^2}+2m^4_{h^{++}}\ln\frac{m^2_{h^{++}}}{Q^2}\right)\crn
&+&\frac{3}{64\pi^2}\left(2m^4_U\ln\frac{m^2_U}{Q^2}+2m^4_V\ln\frac{m^2_V}{Q^2}-
12m^4_Q\ln\frac{m^2_Q}{Q^2}\right)\crn &+&\frac{T^4}{4\pi^2}\left[
F_{-}\left(\frac{m_{h_2}}{T}\right)+2F_{-}\left(\frac{m_{h^{++}}}{T}\right)\right]\crn
&+&\frac{3T^4}{4\pi^2}\left[F_{-}\left(\frac{m_{Z_2}}{T}\right)+2F_{-}\left(\frac{m_U}{T}\right)
+2F_{-}\left(\frac{m_V}{T}\right)+12F_{+}\left(\frac{m_Q}{T}\right)\right],\nn
\eea and  \bea
V^{light}_{eff}&=&\frac{3}{64\pi^2}\left(m^4_{Z_1}\ln\frac{m^2_{Z_1}}{Q^2}
+2m^4_W\ln\frac{m^2_W}{Q^2}-4m^4_t\ln\frac{m^2_t}{Q^2}\right)\crn
&+&\frac{3T^4}{4\pi^2}\left[F_{-}\left(\frac{m_{Z_1}}{T}\right)+
2F_{-}\left(\frac{m_W}{T}\right)+4F_{+}\left(\frac{m_t}{T}\right)\right].\nn
\eea Here $V^{light}_{eff}$ is like the effective potential of the
SM, while  $V^{hard}_{eff}$ is contributions from  heavy
particles. We expect that $V_{hard}^{eff}$  contributes heavily in
the EWPT.

The symmetry breaking in the RM331 can take place sequentially.
Because two scales of symmetry breaking are very different,
$v_{\chi_0} \gg  v_{\rho_0}$ ($v_{\chi_0}\sim 4-5$ TeV,
 $v_{\rho_0}=246$ GeV) and because of the accelerating universe, the symmetry breaking
  $SU(3)\rightarrow SU(2)$ takes place before the symmetry breaking $SU(2)\rightarrow U(1)$.
  The symmetry breaking $SU(3)\rightarrow SU(2)$ through $\chi_0$, generates the
   masses of the heavy gauge bosons such  as $U^{\pm\pm}$, $V^{\pm}$, $Z_2$, and exotic quarks.

Through the  boson mass formulations  in the above sections, we see that boson $V^{\pm}$
 only involves in the phase transition $SU(3) \rightarrow SU(2)$. $Z_1$, $W^{\pm}$
 and $h_1$ only involve in the phase transition $SU(2) \rightarrow U(1)$. However,
 $U^{\pm\pm}$, $Z_2$ and $h^{--}$ involve in both two phase transitions.

The first one is the phase transition $SU(3) \rightarrow SU(2)$.
This phase transition involves exotic quarks, heavy bosons,
without involvement of the SM particles, so $v_{\rho} $ is omitted
in this phase transition.  The effective potential can be rewritten
 as  follows \cite{cos2}
 \[ V_{SU(3) \rightarrow SU(2)}^{eff}=D'(T^2-{T'}^2_0){v_{\chi}}^2-E'
Tv_{\chi}^3 +\frac{\lambda'_T}{4}v_{\chi}^4
\] The minimum  conditions are \[ V_{eff}(\chi_0)=0;\hs
V'_{eff}(\chi_0)=0;\hs  V''_{eff}(\chi_0)=m^2_{h_2}, \] where
\bea
D'&=& \frac{1}{24 {v_{\chi_0}}^2} \left\{6 m_U^2+ 3m_{Z_2}^2+6 m_{V}^2
+ 18m_Q^2+ 2 m_{h^\pm}^2 \right\},\crn
{T'}_0^2 &=&  \frac{1}{D}\left\{\frac{1}{4} m_{h_2}^2 -
 \frac{1}{32\pi^2v_{\chi_0}^2} \left(6 m_U^4+ 3 m_{Z_2}^4+6 m_{V}^4 -
 36 m_Q^4 + 2 m_{h^\pm}^4\right)\right\},\crn
E' &=& \frac{1}{12 \pi v_{\chi_0}^3} (6 m_U^3 + 3 m_{Z_2}^3 +
6 m_{V}^3 + 2 m_{h^\pm}^3),\crn
\lambda'_T &=&
 \frac{m_{h_2}^2}{2 v_{\chi_0}^2} \left\{ 1 - \frac{1}{8\pi^2 v_{\chi_0}^2 m_{h_2}^2}
 \left[6 m_V^4 \ln \frac{m_V^2}{b T^2} +3 m_{Z_2}^4 \ln \frac{m_{Z_2}^2}{b T^2}
 \right.\right.\crn
&&\qquad \left.\left. +6 m_U^4 \ln \frac{m_U^2}{bT^2}-36 m_Q^4 \ln
\frac{m_Q^2}{b_F T^2} + 2  m_{h^\pm}^4 \ln \frac{m_{h^\pm}^2}{b
T^2} \right] \right\}.\nn
\eea
The critical temperature is determined  as follows \be
T'_c=\frac{T'_0}{\sqrt{1-E'^2/D'\lambda'_{T'_c}}}.\label{th} \ee

For simplicity, let us   assume $m_{h_2}=X$,
$m_{h^{--}}=m_{Z_2}=m_{Q}=K$. In order to have the first-order phase
transition, the phase transition strength  must be larger than 1,
i.e., $\frac{v_{\chi_c}}{T'_c} \ge 1$.

If $X$ is larger than $200$ GeV,  the heavy particle
masses  are in range of few TeVs in order to have the first-order phase transition \cite{cos2}.
 In order to have the first-order phase transition, if the contribution
of $h_2$ with the mass is smaller than 200 GeV,  $K$ is smaller than 1.5 TeV \cite{cos2}.

 The second/last step is the phase transition $SU(2) \rightarrow U(1)$.
This phase transition does not involve the exotic quarks and boson $V^{\pm}$.
Hence, in this case,  $v_{\chi} $ is neglected, and  the contribution of $U^{\mp\mp}$ is equal to
 $ W^{\mp}$. Then
\bea
V_{SU(2) \rightarrow U(1)}^{eff}&= &
v_0(v_\rho)\frac{1}{64\pi^2}\left(m^4_{h_2}\ln
\frac{m^2_{h_2}}{Q^2}+2m^4_{h^{++}}\ln\frac{m^2_{h^{++}}}{Q^2}\right)\crn
&+&\frac{3}{64\pi^2}\left(2m^4_U\ln\frac{m^2_U}{Q^2}+m^4_{Z_1}
\ln\frac{m^2_{Z_1}}{Q^2}\right. \crn
&+ & \left. m^4_{Z_2}\ln\frac{m^2_{Z_2}}{Q^2}
+2m^4_W\ln\frac{m^2_W}{Q^2}-4m^4_t\ln\frac{m^2_t}{Q^2}\right)\crn
&+&\frac{T^4}{4\pi^2}\left[F_{-}\left(\frac{m_{h_2}}{T}\right)+
2F_{-}\left(\frac{m_{h^{++}}}{T}\right)\right]\crn
&+&\frac{3T^4}{4\pi^2}\left[2F_{-}\left(\frac{m_U}{T}\right)+F_{-}\left(\frac{m_{Z_1}}{T}\right)
\right.\crn
&+& \left.F_{-}\left(\frac{m_{Z_2}}{T}\right)+2F_{-}\left(\frac{m_W}{T}\right)
+4F_{+}\left(\frac{m_t}{T}\right)\right]
\nn\eea

Denoting $V_{SU(2) \rightarrow U(1)}^{eff} \equiv  V_{SU(2) \rightarrow U(1)}^{eff}(v_{\rho}, T)$,
at  high-temperature, it becomes
\[ V_{eff}^{RM331}=D(T^2-T^2_0).v^2_{\rho}-ET|v_{\rho}|^3+\frac{\lambda_T}{4}v^4_{\rho},
\]
where
\bea
D &=& \frac{1}{24 {v_0}^2} \left\{6 m_W^2+6 m_U^2+ 3 m_{Z_1}^2
+3 m_{Z_2}^2+ 6 m_t^2+m_{h_2}^2 + 2 m_{h^\pm}^2 \right\},\crn
T_0^2 &=&  \frac{1}{D}\left\{\frac{1}{4} m_{h_1}^2 -
 \frac{1}{32\pi^2v_0^2} \left(6 m_W^4+6 m_U^4+ 3 m_{Z_1}^4
 +3 m_{Z_2}^4 - 12 m_t^4 \right.\right.\crn
&&\qquad \left.\left.
+ m_{h_2}^4 + 2 m_{h^\pm}^4\right)
\right\},\crn
E &=& \frac{1}{12 \pi v_0^3} (6 m_W^3+6 m_U^3
+ 3 m_{Z_1}^3 +3 m_{Z_2}^3 + m_{h_2}^3 + 2 m_{h^\pm}^3),\label{trilinear}\\
\lambda_T &=&
 \frac{m_{h_1}^2}{2 v_0^2}
 \left\{ 1
- \frac{1}{8\pi^2 v_0^2 m_h^2}
 \left[6 m_W^4 \ln \frac{m_W^2}{b T^2} + 3 m_{Z_1}^4 \ln
 \frac{m_{Z_1}^2}{b T^2}+3 m_{Z_2}^4 \ln \frac{m_{Z_2}^2}{b T^2}\right.\right.\crn
&&\qquad \left.\left.
+6 m_U^4 \ln \frac{m_U^2}{bT^2}-12 m_t^4 \ln \frac{m_t^2}{b_F T^2}
+  m_{h_2}^4 \ln \frac{m_{h_2}^2}{b T^2} + 2  m_{h^\pm}^4 \ln \frac{m_{h^\pm}^2}{b T^2}
\right]
\right\},\nn
\eea
here we have  assumed $m_{H_2}=m_{ h_{--}}=m_{Z_2}\equiv Y$  with boson $Z_2$  and used
 $Q \equiv v_{\rho_0}=v_0=246$ GeV.

 In order to have the first-order
 phase transition, the phase transition strength has to be  larger than 1, i.e.,
 $\frac{v_{\rho_c}}{T_c}\geq 1$. The critical temperature $T_c$ is given by
\be T_c=\frac{T_0}{\sqrt{1-E^2/D\lambda_{T_c}}}.\label{th}
\ee

To survive the critical temperatures,  $T_c$, $T_0$ must be positive, so $T_0$ is
also positive, from which we can draw on conditions for heavy particles. Therefore, we get
\[ \frac{1}{4} m_{h_1}^2 -  \frac{1}{32\pi^2v_0^2} \left(6 m_W^4+6 m_U^4+ 3 m_{Z_1}^4
+3 m_{Z_2}^4 - 12 m_t^4 + m_{h_2}^4 + 2 m_{h^\pm}^4\right) >0
\]
With $m_{h_1}=125$ GeV and  assuming $m_{Z_2}=m_{h_2}=m_{h^{--}}=Y$,
we can obtain $Y<344.718$ GeV \cite{cos2}.

When $\frac{v_{\rho_c}}{T_c} = 1$, i.e., $2E/\lambda_{T_c}=1$, we
obtain $Y=203.825$ GeV, and the critical temperature
is in range $0 < T_c < 111.473$ GeV.
The contributions of new particles make of the strongly first-order phase transition
 that the SM cannot. However, there is one thing special, heavy
  particles as $U^{\pm\pm}, h_2, h_ {--}, Z_2$ that contribute only the little part in their mass.

 When temperature goes  close to $ T_c $, the second
 minimum slowly formed distinct, i.e.,  the phase transition nucleation appears.

When temperature goes  over $T_c$, the minimum
goes  to zero, i. e.,   the symmetry phase is restored.
This was showed that phase transition
$SU(2) \rightarrow U(1)$ is the first-order phase transition \cite{cos2}.

We find that the effective potential of this model is different from that of the SM, and  it has
contributions from heavy bosons as triggers for the strongly first-order phase transition
with $m_{h_1}=125$ GeV.

We have got the  following  constraints on the mass of Higgs in RM331 \cite{cos2}
\[
285.56\, {\rm GeV}<M_{h_2}< 1.746 \, {\rm TeV},\hs
3.32\,  {\rm TeV} <M_{h_{--}}< 5.61\,  {\rm TeV}.
\]

Thus we have used the effective potential at finite temperature to study the structure of the
EWPT in the RM331 model. This phase transition is split into
two phases, namely, the first transition  is $SU (3)\rightarrow SU(2)$ or the symmetry
breaking in the energy scale $ v_ {\chi_0}$ in order to generate masses for heavy
 particles and exotic quarks. The second phase transition is
 $ SU (2)\rightarrow U(1)$  at $ v_ {\rho_0} $.  The EWPT in this model may be
  the strongly first-order EWPT with $m_{h_1}= 125$ GeV if the heavy bosons masses  are some
  few TeVs.

\subsection{Phase transition in economical 3-3-1 model}
 In this section, we follow the same approach for E331 model \cite{ecn331}, whose lepton sector is more
 complicated than that of the RM331 model. The E331 model has the right-handed neutrino in the leptonic content,
 the bileptons (two singly charged gauge bosons $W^\pm$, $Y^\pm$, and a neutral  gauge bosons $X^0$), the heavy
 neutral  boson $Z_2$, and the exotic quarks. The masses of particles in the E331
 were summarized in Table \ref{tab5}
 \begin{table}
\caption{Mass formulations of bosons in the E331 model}
\bc
\begin{tabular}{|l|l|c|c|}
\hline
\hline
 Bosons& $m^2(\om,v)$& $ m^2(\om)$& $m^2(v)$    \\
\hline
$m_{W^{\pm }}^{2}$& $\frac{g^{2}}{4}v^{2}$&
0& $80.39^2$ $(\mathrm{GeV})^2$ \\
\hline
$m_{Y^{\pm }}^{2}$& $\frac{g^{2}}{4}(\om^2+v^{2})$&
$\frac{g^{2}}{4}\om^2$& $80.39^2$ $(\mathrm{GeV})^2$\\
\hline
$m_{X^0}^{2}$& $\frac{g^{2}}{4}\om^2$&
$\frac{g^{2}}{4}\om^2$& 0 \\
\hline
$m^2_{Z_1}\sim m^2_{Z}$& $\frac{g^2}{4c_W^2} v^2$&
0& $91.68^2$ $(\mathrm{GeV})^2$ \\
\hline
$m^2_{Z_2}\sim m^2_{Z'}$& $\frac{g^2c_W^2}{3-4s_W^2}\om^2$
&$\frac{g^2c_W^2}{3-4s_W^2}\om^2$& 0 \\
\hline
$m^2_{H^0}$& $\left(2\la_2 -\frac{\la^2_3}{2\la_1}\right)v^2$&
0& $125^2$ $(\mathrm{GeV})^2$\\
\hline
$m^2_{H^0_1}$& $2\lambda_1\om^2+\frac{\la^2_3}{2\la_1}v^2$ &
$2\lambda_1\om^2$& $\frac{\la^2_3}{2\la_1}v^2$\\
\hline
$m^2_{H^\pm_2}$& $\fr{\la_4}{2}(\om^2+v^2)$&
$\fr{\la_4}{2}\om^2$ & $\fr{\la_4}{2}v^2$ \\
\hline
\hline
\end{tabular}
\ec
\label{tab5}
\end{table}

As in the RM331, here  EWPT takes place with two  transitions: i)  $SU(3) \rightarrow SU(2)$ at the
scale of $\om_0$ and the transition $SU(2)\rightarrow U(1)$ at the scale of  $v_0$ \cite{cos4}.

The first phase transition $SU(3)\rightarrow SU(2)$ due to $\om$ provides the bounds on
 parameters presented in Table \ref{tab6}

\begin{table}
\caption{The mass ranges of $H^0_1$ and $H^{\pm}_2$ for the first-order EWPT $SU(3) \rightarrow SU(2)$
 and their upper bounds  by the condition $m_{boson}< 2.2 \times {T'_c}$.}
\bc
\begin{tabular}{|c|c|c|c|c|}
\hline
\hline
$\om \, [TeV]$ & $T'_c \, [GeV] $ & ${m_{H^0_1}} \, [GeV]$ & ${m_{H^{\pm}_2}} \, [GeV]$ &
Upper bound $[GeV]$ \\
\hline
$1$     & $350$   & $0<m_{H^0_1}<300$ & $0<m_{H^{\pm}_2}<720$ &  $770$ \\
\hline
$2$     & $650$   & $0<m_{H^0_1}<600$ & $0<m_{H^{\pm}_2}<1440$ &  $1430$ \\
\hline
$3$     & $950$   & $0<m_{H^0_1}<900$ & $0<m_{H^{\pm}_2}<2150$ &  $2090$ \\
\hline
$4$     & $1300$   & $0<m_{H^0_1}<1200$ & $0<m_{H^{\pm}_2}<2870$ &  $2860$ \\
\hline
$5$     & $1600$   & $0<m_{H^0_1}<1500$ & $0<m_{H^{\pm}_2}<3590$ &  $3520$ \\
\hline
\hline
\end{tabular}
\ec
\label{tab6}
\end{table}

The new bosons and exotic
  quarks can be triggers for the EWPT $SU(3) \rightarrow SU(2)$ to be the first-order.
It was shown that the EWPT $SU(2) \rightarrow U(1)$ is the first-order phase transition, but it seems quite weak \cite{cos4}.

\section{\label{spha} Electroweak sphalerons in the reduced minimal 3-3-1 model}

To be consistent with cosmological evolution, our strategy is the following: the model has to have an inflation or
 phase transition of the first-order. As a result, the leptogenesis or CP-violation exist. Then sphaleron completes to
produce the BAU.  Sphaleron is a transition at high temperature where  thermal fluctuations can bring the
 magnitude of the Higgs field from zero VEV over the barrier to nonzero VEV classically without tunneling.
 In \cite{cos3}, the sphalerons in the RM331 were considered.
In the SM, the sphaleron rate is very small, about $10^{-60}$ \cite{manton, sphasm, Farrar, Arnold, spha-nonpur};
this rate is much smaller than the rate of BAU and smaller than the cosmological expansion rate.

To study the sphaleron processes, we consider the Lagrangian of the gauge-
Higgs system
\be\label{gauge-Higgs}
\mathcal{L}_{\rm gauge-Higgs} =-\fr1 4 F^{a}_{\mu\nu}F^{a\mu\nu}
    + \left( \mathcal{D}_{\mu }\chi \right) ^{\dagger }\left( \mathcal{%
D}^{\mu }\chi \right) +\left( \mathcal{D}_{\mu }\rho \right)
^{\dagger }\left( \mathcal{D}^{\mu }\rho \right)-V(\chi,\rho).
\ee
 Assuming  the least energy has the pure-gauge
configurations  ($F^a_{ij}=0$), we get   energy functional in the temporal gauge
\be \label{Energy-01}
\mathcal{E} = \int d^3 \textbf{x}
    \left[\left( \mathcal{D}_{\mu }\chi \right) ^{\dagger }\left( \mathcal{%
D}^{\mu }\chi \right) +\left( \mathcal{D}_{\mu }\rho \right)
^{\dagger }\left( \mathcal{D}^{\mu }\rho
\right)+V(\chi,\rho)\right],
\ee

By the temperature expansion, the energy functional is given by
\be\label{Energy-02}
\mathcal{E} = 4\pi\int^\infty_0d^3\textbf{x}\left[\frac{1}{2}\left(\nabla^2 v_\chi\right)^2
+\fr 1 2 \left(\nabla^2 v_\rho\right)^2
    +V_{\rm eff}(v_\chi, v_\rho;T)\right].
\ee

In  the static field approximation, we have two equations of motion for the VEVs in spherical coordinates \cite{cos3}
for the VEVs:
\be\label{cda}
\ddot{v_\chi}+\nabla^2 v_\chi-\frac{\partial V_{eff}(v_\chi,T)}{\partial v_\chi} = 0,
 \ee
and
 \be \label{cdb}
\ddot{v_\rho}+\nabla^2 v_\rho-\frac{\partial V_{eff}(v_\rho,T)}{\partial v_\rho}=0.
\ee
Then, the sphaleron energies in the $SU(3)\rightarrow SU(2)$
 and $SU(2)\rightarrow U(1)$ phase transitions, are given,  respectively
\be\label{4.163a}
 \mathcal{E}_{sph.su(3)}=4\pi\int\left[\frac{1}{2}\left(\frac{dv_\chi}{dr}\right)^2+V_{eff}(v_\chi,T)\right]r^2dr,
 \ee
and
 \be \label{4.163b}
\mathcal{E}_{sph.su(2)}=4\pi\int\left[\frac{1}{2}\left(\frac{dv_\rho}{dr}\right)^2+V_{eff}(v_\rho,T)\right]r^2dr.
\ee

The sphaleron rate per unit time per unit volume,
$\Gamma/V$, is characterized by a Boltzmann factor,
$\exp\left(-\mathcal{E}/T\right)$, as follows \cite{Arnold, ctspha,gauge0}:
\be \label{SphaleRateDef}
\Gamma/V = \alpha^4 T^4 \exp\left(-\mathcal{E}/T\right),
\ee
where $V$ is the volume of the EWPT's region,  $T$ is the
temperature, $\mathcal{E}$ is the sphaleron energy, and
$\alpha=1/30$.

We will compare the sphaleron rate with the Hubble
constant, which describes the cosmological expansion rate at the
temperature $T$ \cite{cross, spha-huble}
\be \label{Hubble}
H^2=\frac{\pi^2 g T^4}{90 M_{pl}^2},
\ee
where  $g=106.75$, $M_{pl}=2.43 \times 10^{18}$ GeV.

 Assuming that the VEVs of the Higgs fields do not change from point to point in the universe,
 then  we have $\frac{dv_\chi}{dr}=\frac{dv_\rho}{dr}=0$, and
\be\label{dkk}
\frac{\partial V_{eff}(v_\chi)}{\partial v_\chi}=0, \quad \frac{\partial V_{eff}(v_\rho)}{\partial v_\rho}=0.
\ee

Eqs. (\ref{dkk}) shows that $v_\chi$ and $v_\rho$ are the
extremes of the effective potentials. The sphaleron energies  can be rewritten as
\be\label{4.165a}
 \mathcal{E}_{sph.su(3)}=4\pi\int V_{eff}(v_\chi,T) r^2dr=\frac{4\pi r^3}{3}V_{eff}(v_\chi,T) \bigg|_{v_{\chi_m}},
 \ee
and
 \be \label{4.165b}
 \mathcal{E}_{sph.su(2)}=4\pi\int V_{eff}(v_\rho,T) r^2dr=\frac{4\pi r^3}{3}V_{eff}(v_\rho,T) \bigg|_{v_{\rho_m}},
 \ee
where $v_{\chi_m}, v_{\rho_m}$ are the VEVs at the maximum of the effective potentials. From  (\ref{4.165a}) and
 (\ref{4.165b}), it follows that  the sphaleron
 energies are equal to the maximum heights of the potential barriers.

The universe's volume  at a temperature $T$ is given by $V=\frac{4\pi r^3}{3}=\frac{1}{T^3}$. Because the whole
 universe is an identically thermal bath, the sphaleron energies are approximately
\be\label{4.20}
 \mathcal{E}_{sph.su(3)}\sim \frac{{E'}^4 T}{4{\lambda'}_T^3}; \quad \mathcal{E}_{sph.su(2)}\sim \frac{E^4 T}{4\lambda^3_{T}}.
\ee
From the definitions ( \ref{4.165a}) and (\ref{4.165b}), the sphaleron rates take the form, respectively
\be\label{4.21a}
\Gamma_{su(3)}=\alpha_w^4 T\exp\left({-\frac{{E'}^4T}{4{\lambda'}_{T}^3T}}\right) ,
 \ee
and
 \be \label{4.21b}
\Gamma_{su(2)}=\alpha_w^4 T\exp\left({-\frac{E^4T}{4\lambda_{T}^3T}}\right).
 \ee

For the heavy particles, $E, \lambda, E'$ and $\lambda'$ are constant, and  the sphaleron rates (for the
 the phase transition $SU(2)\rightarrow U(1)$) in
 this approximation are the linear functions of temperature \cite{cos3}

Thus,  the upper bounds of the sphaleron rates are much larger the Hubble constant \cite{cos3}
\be\label{UpBound}
\Gamma_{su(3)}\sim 10^{-3}\gg H; \quad \Gamma_{su(2)} \sim 10^{-4} \gg  H \sim 10^{-13}.
\ee

In a thin-wall approximation, sphaleron rates are presented in Tables  \ref{tab.02} and \ref{tab.03}

\begin{table}
\caption{The sphaleron rate in the EWPT $SU(3)\rightarrow SU(2)$ with $m_q(v_\chi)=m_{h2}(v_\chi)=1500\,  {\rm GeV}$.}
\begin{tabular}{l|l|c|c|c|c|c}
\hline\hline
$T$              & $R_{b.su(3)}$  & $R_{b.su(3)}/\Delta l'$  & $\mathcal{E}_{sph.SU(3)}$ & $\Gamma_{SU(3)}$& $H$   & $\Gamma_{SU(3)}/H$         \\
$[GeV]$ &$[10^{-6}\times GeV^{-1}]$ & &$[GeV]$ &$[10^{-11}\times GeV]$ & $[10^{-12}\times GeV]$ &\\
\hline
1479.48 ($T'_1$)  &$10$      &$10$         &$6975.17$            &$1.63719 \times 10^{6}$ &$3.08195$ &$5.31 \times 10^6$\\
\hline
1450                      &$12$       &$12$         &$12481.3$           &$3.2702 \times 10^{4}$    &$2.96034$ &$1.10 \times 10^5$\\
\hline
1400                      &$13$       &$13$         &$17206.3$           &$7.94481 \times 10^2$      & $2.7597$ &$2.878 \times 10^3$\\
\hline
1390                      &$15$      & $15$         & $23251.7$           &$9.3264$   &$2.72042$ & $3.42$ \\
\hline
1388.4556 ($T'_c$) &$16.5$  & $16.5$       & $28135.1$          & $0.2714$   &$2.71438$  & 1 \\
\hline
1387                      &$17$      & $17$         & $29854.0$          & $0.07687$  & $2.70869$ & $0.28$ \\
\hline
1000                     &$19$       &$19$          & $60590.8$          &$5.98\times 10^{-19}$ &$1.40801$ &$4.25 \times 10^{-18}$\\
\hline
900                      & $22$       &$22$          & $89250.8$         &$9.50\times 10^{-36}$ & $1.14049$ & $8.33 \times 10^{-35}$\\
\hline
865.024 ($T'_0$)    &$25$        &$25$          &$119110.36$      &$1.69 \times 10^{-52}$ & $1.05357$ & $1.60 \times 10^{-51}$\\
\hline\hline
\end{tabular}\label{tab.02}
\end{table}

\begin{table}
\caption{The sphaleron rate in the EWPT $SU(2)\rightarrow U(1)$ with $m_{h2}(v_\rho)=100 GeV, m_{h^{\pm\pm}}(v_\rho)=350 GeV$.}
\begin{tabular}{l|l|c|c|c|c|c}
\hline\hline
$T$                      & $R_{s.su(2)}$  & $ R_{s.su(2)}/\Delta l$  & $\mathcal{E}_{sph.SU(2)}$ & $\Gamma_{SU(2)}$ & $H$ & $\Gamma_{SU(2)}/H$                                           \\
$[GeV]$ &$[10^{-4}\times GeV^{-1}]$ & &$[GeV]$ &$[10^{-12}\times GeV]$ & $[10^{-14}\times GeV]$ &\\
\hline
141.574 ($T_1$) &$6$              & $10$          & $742.838$       &$919936.07$         & $2.82211$  &$3.25\times 10^{7}$\\
\hline
141.5                   &$8$              & $10$          & $1020.87$       & $128525.28$         & $2.81916$  &$4.55 \times 10^6$\\
\hline
141                      &$10$            &$10$           &$1442.75$        &$6264.89$             & $2.79927$   & $2.23 \times 10^5$\\
\hline
140                      & $12$           &  $12$         & $2342.21$        &$9.37289$             &$2.7597$     & $339.6$\\
\hline
138.562 ($T_c$)  &$13.1$          & $13$         & $3135.75$        & $0.02703$            &$2.703$       &1    \\
\hline
137                      &$14$            & $14$         & $3922.29$         & $0.0000622$         & $2.6427$    & $2.357 \times 10^{-3} $\\
\hline
130                      &$16$          & $16$          &$6567.08$          &$1.847 \times 10^{-14}$ & $2.379$ & $7.76 \times 10^{-13}$\\
\hline
120                      & $18$         &$18$          &$10068.2$          &$5.403 \times 10^{-29}$ &$2.02754$ &$2.66 \times 10^{-27}$\\
\hline
118.42 ($T_0$)    &$20$          &$20$          &$12656.7$           &$5.595 \times 10^{-39}$ &$6.209$ &$9.01 \times 10^{-38}$\\
\hline\hline
\end{tabular}\label{tab.03}
\end{table}
Here  $R_{b.su(3)}$ and $\Delta l'$ are respectively the radius and the wall thickness of a bubble which is nucleated in the phase transitions.

We conclude that the sphaleron rates are larger than the cosmological expansion rate at temperatures above the critical temperature and are smaller than
   the cosmological expansion rate at temperatures below the critical temperature.
    For each transition, baryon violation rapidly takes place in the symmetric phase regions but it also quickly shuts
     off in the broken phase regions. This may provide B-violation necessary for baryogenesis, as required by the first of Sakharov's conditions, in the connection with non-equilibrium physics.

 \section{\label{conclus}Conclusions}
In this review, we have showed that the 3-3-1 models are able to describe the cosmological
evolution. The 3-3-1 models contain the hybrid inflationary scenario and the first-order phase transitions.
The inflation happens in the GUT scale, while  phase transition has two sequences corresponding two steps
of symmetry breaking in the models.
The sphaleron rates are much larger than the Hubble constant. They are larger than the cosmological expansion
 rate at temperatures above the critical temperature and are smaller than the cosmological expansion rate at temperatures below the critical temperature.
From these considerations, some bound on model parameters are deduced.

\section*{Acknowledgments} The author thanks  D. T. Huong, V. Q. Phong, V. T. Van for collaboration.
This research is funded by Vietnam National Foundation for Science
and Technology Development (NAFOSTED) under grant number 103.01-2014.51.


\begin{thebibliography}{99}

\bibitem{plank}P. A. R. Ade \emph{ et al},  (Planck Collab. 2013 XXIV), arXiv:1303.5084v1.


\bibitem{sakharov} A.D. Sakharov, JETP Lett.\textbf{5}, 24 (1967).
\bibitem{mkn} V.  Mukhanov,  {\it Physical Foundations of Cosmology},
 (Cambridge University Press, Cambridge, England, 2005).

\bibitem{ppf} F. Pisano and V. Pleitez, Phys. Rev. D \textbf{46}, 410 (1992); P. H. Frampton, Phys. Rev. Lett.  \textbf{69}, 2889 (1992);
 R. Foot {\it et al,} Phys. Rev. D \textbf{47}, 4158 (1993).

\bibitem{flt} M. Singer, J. W. F. Valle, and J. Schechter, Phys. Rev. D \textbf{22}, 738 (1980);  R. Foot, H. N. Long, and Tuan A. Tran,
 Phys. Rev. D \textbf{50} R34 (1994),  [arXiv:hep-ph/9402243];
 J. C. Montero, F. Pisano, and  V. Pleitez, Phys. Rev. D \textbf{47},
 2918 (1993); H. N. Long, Phys. Rev. D \textbf{53}, 437 (1996); H. N. Long, Phys. Rev. D \textbf{54}, 4691 (1996) ;
  H.  N. Long, Mod. Phys. Lett. {\bf A13}, 1865 (1998).


\bibitem{chargequan} C. A. de S. Pires, O. P. Ravinez, Phys. Rev. D
\textbf{58}, 035008 (1998); A. Doff, F. Pisano,  Mod. Phys.
Lett. A \textbf{14}, 1133 (1999); Phys. Rev. D \textbf{63}, 097903 (2001); P. V. Dong,
H. N. Long,  Int. J. Mod. Phys. A \textbf{21}, 6677 (2006).


\bibitem{uni} S. M. Boucenna, R. M. Fonseca, F. Gonlzalez-Canales, and J. W. F. Valle,
Phys. Rev. D \textbf{91}, 031702 (2015),
arXiv:1411.0566[hep-ph].



\bibitem{changlong} D. Chang and H. N. Long, {\it Phys. Rev}. D {\bf 73}, (2006) 053006,
[arXiv: hep-ph/0603098]; see also, M. B.  Tully and G. C. Joshi, \emph{Phys. Rev}. D \textbf{64}, 011301 (2001).


\bibitem{longvan} H. N. Long and  V. T. Van, {\it J. Phys.} G: Nucl. Part. Phys. {\bf 25},
(1999) 2319,  [arXiv: hep-ph/9909302]; J. M. Cabarcas, D. Gomez Dumm, R. Martinez,  {\it J. Phys.}
 G 37 (2010) 045001;
A. C. B. Machado, J. C. Montero and V. Pleitez,  Phys. Rev. D {\bf 88} (2013) 113002,
arXiv:1305.1921 [hep-ph].

\bibitem{331LHC}J. E. Cieza Montalvo \emph{ et al,} 	
{\it Neutral 3-3-1 Higgs Boson Through $e^{+} e^{-}$ Collisions},
 arXiv:1408.5944 [hep-ph]; J .E. Cieza Montalvo  \emph{et a}l, Phys. Rev. D \textbf{88} (2013), 095020;
J. E. Cieza Montalvo \emph{et a}l,  Nucl. Phys. \textbf{A790} (2007) 554; 	
J.  E. Cieza Montalvo\emph{ et al}, Phys. Rev. D \textbf{76} (2007) 117703.

\bibitem{331ep}J .E. Cieza Montalvo\emph{ et al},  Eur. Phys. J. \textbf{C 72} (2012) 2210;  	J. E. Cieza Montalvo \emph{et al},
 Phys. Rev. D \textbf{78} (2008) 116003.

\bibitem{Higgsdecay331}
C-X. Yue, Q-Y. Shi, T.  Hua, Nucl. Phys. \textbf{B 876} (2013) 747,
arXiv:1307.5572 [hep-ph].

\bibitem{cataenoEPJC} W. Caetano, C. A. de S. Pires, P. S. Rodrigues da Silva, D. Cogollo,
F. S. Queiroz,  Eur. Phys. J. \textbf{C 73} (2013) 2607.


\bibitem{g2muon} N. A.  Ky, H. N. Long, D. V. Soa, Phys. Lett. B 486 (2000) 140, [hep-ph/0007010];
C. A. de S. Pires, P. S. Rodrigues da Silva, Phys. Rev. D 64 (2001) 117701, [hep-ph/0103083];
 C. A. De S. Pires, P. S. Rodrigues da Silva, Phys. Rev. D 65 (2002) 076011, [hep-ph/0108200];
C.  Kelso, P. R. D. Pinheiro, F.  S. Queiroz, W.  Shepherd. Eur. Phys. J. C 74 (2014) 2808;
Chris Kelso, H. N. Long, R. Martinez, Farinaldo S. Queiroz,
 Phys. Rev.  \textbf{D 90}, 113011 (2014)
[arXiv:1408.6203(hep-ph)].


\bibitem{neutrino331}
 P. V. Dong, H. N. Long, D. V. Soa, and V. V. Vien,   \emph{Eur.
Phys. J. C} \textbf{71}, 1544 (2011), arXiv:1009.2328 [hep-ph];
P. V. Dong, H. N. Long, C. H. Nam, and V. V. Vien,  \emph{Phys. Rev.
 D} \textbf{85}, 053001(2012), arXiv: 1111.6360 [hep-ph]; V. V. Vien and H. N. Long,
JHEP \textbf{04} (2014)133, [arXiv:1402.1256 (hep-ph)];  V. V. Vien and  H. N. Long,
 {\it Int. J. Mod. Phys.}  {\bf A  28} (2013), No. 32,
1350159, [arXiv:1312.5034(hep-ph)]; S. M. Boucenna, S. Morisi and J. W. F. Valle,
Phys. Rev. D \textbf{90}, 013006 (2014); for a review, see C. A. de S. Pires,\
 \emph{Physics International} (PI)  \textbf{6}, No 1 (2015) pp.  33-41,
 arXiv:1412.1002 [hep-ph].

\bibitem{dm331} D. Fregolente and M. D. Tonasse, Phys. Lett. \textbf{B 555}, 7 (2003); H. N. Long and
N. Q. Lan, Europhys. Lett. \textbf{64}, 571 (2003);  N. Q. Lan and H. N. Long,
{\it  Astrophys. Space Sci}.  {\bf 305}  (2006) 225;
S. Filippi, W. A. Ponce and L. A. Sanches,
 Europhys. Lett. \textbf{73}, 142 (2006);
 C. A. de S. Pires and P. S. Rodrigues da Silva, JCAP \textbf{12} (2007) 012;
  K. Mizukoshi, C. A. de S. Pires, F. S. Queiroz, P. S. Rodrigues da Silva, Phys. Rev. D \textbf{83} (2011) 065024;
  J. D. R. Alvares, C. A. de S. Pires, Farinaldo S. Queiroz, D. Restrepo, P. S. Rodrigues da Silva, Phys. Rev. D \textbf{86}, 075011 (2012);
  S.  Profumo, Farinaldo S. Queiroz, Eur. Phys. J. C \textbf{74} (2014)2960;
  Chris Kelso, C. A. de S. Pires, S. Profumo, Farinaldo S. Queiroz, P. S. Rodrigues da Silva, Eur. Phys. J. C \textbf{74} (2014) 2797;
  D. Cogollo, Alma X. Gonzalez-Morales, Farinaldo S. Queiroz, P. R. Teles, JCAP \textbf{11} (2014) 002;
P. V. Dong, T. D. Tham, H. T. Hung, Phys. Rev.  D {\bf 87}, 115003 (2013);  P. V. Dong, D. T. Huong, F.  S. Queiroz, and  N. T. Thuy, Phys. Rev.  D {\bf 90}, 075021 (2014);  for a review see P. S. Rodrigues da Silva, \emph{Physics International} (PI)  \textbf{7}, No 1 (2016) pp.  15-27, [arXiv: 1412.8633(hep-ph)].


\bibitem{s331} P. V. Dong, N. T. K. Ngan and D. V. Soa, {\it Phys. Rev}. D {\bf 90}, (2014) 075019.

\bibitem{dias}
A. G. Dias, R. Martinez, V. Pleitez,  Eur. Phys. J. C {\bf 39}
(2005), 101. See also,  A. G. Dias, V. Pleitez, Phys. Rev. D {\bf
80}, 056007  (2009).



\bibitem{rm331} J. G. Ferreira, Jr., P. R. D. Pinheiro, C. A. de S. Pires, and P. S. Rodrigues da Silva, Phys. Rev. D {\bf 84},
 095019 (2011); V. T. N. Huyen, T. T. Lam, H. N. Long, and V. Q. Phong, Commun.  in Phys. \textbf{24},
  No. 2,  97 (2014), [arXiv:1210.5833(hep-ph)].

\bibitem{dsi} P. V. Dong and D. T. Si,  {\it Phys. Rev}. D {\bf 90}, (2014) 117703,
[arXiv:1411.4400(hep-ph)].


\bibitem{longinami} H.  N.  Long and T. Inami,  {\it
Phys. Rev}.  D {\bf 61}, (2000) 075002, [arXiv: hep-ph/9902475].



 \bibitem{ecn331} W. A. Ponce, Y. Giraldo and L. A. Sanchez, Phys. Rev. D {\bf 67}, 075001 (2003); P. V.
Dong, H. N. Long, D. T. Nhung and D. V. Soa, Phys. Rev. D {\bf
73}, 035004 (2006); P. V. Dong and H. N. Long, Adv. High Energy
Phys. {\bf 2008}, 739492 (2008), [arXiv:0804.3239(hep-ph)]; P. V. Dong, Tr. T. Huong, D. T. Huong, and H. N. Long, Phys. Rev. D {\bf
74}, 053003 (2006); P. V. Dong, H. N. Long, and D. V. Soa, Phys. Rev. D {\bf 73}, 075005 (2006); P. V. Dong, H. N. Long,
 and D. V. Soa, Phys. Rev. D {\bf 75}, 073006 (2007);  P. V. Dong, H. T. Hung, and H. N. Long, Phys. Rev. D {\bf 86}, 033002 (2012).



\bibitem{infsce} See, for example, A. H. Guth, Phys. Rev. D {\bf 23}, 347
(1981); A.  Linde, Phys. Lett. {\bf 108 B}, 389 (1982); A. Albrecht
and P. J. Steinhardt, Phys. Rev. Lett. {\bf 48}, 1220 (1982).


\bibitem{lind383}  A. Linde, Phys. Lett. {\bf 129  B}, 177 (1983).


\bibitem{cos1} Do T. Huong and Hoang  N. Long, {\it Phys. Atom. Nucl.} {\bf 73}, 791
(2010). [arXiv: 0807.2346].




\bibitem{msusy} J. C. Montero, V. Pleitez, and M. C. Rodriguez,
 Phys. Rev. {\bf D 65}, 035006 (2002).

\bibitem{duongma}  T. V. Duong and E. Ma,  Phys. Lett. {\bf B 316}, 307 (1993);
 M. C. Rodriguez,  Int. J. Mod.
Phys. {\bf  A 21}, 4303 (2006).

\bibitem{leptonmassm331}J. C. Montero, V. Pleitez, and  M. C.
Rodriguez,  Phys. Rev. D {\bf  65},  095008 (2002); C. M. Maekawa
and M. C. Rodriguez,   JHEP \,  {\bf 04}, 031 (2006).


\bibitem{consm331}  M. Capdequi-Peyranere, M. C. Rodriguez,
  Phys. Rev. D {\bf  65},  035001 (2002).


\bibitem{longpal}  H.  N. Long and P. B.  Pal,
  Mod. Phys. Lett. {\bf A 13}, 2355 (1998).

\bibitem{srm331}
D. T. Huong, L. T.  Hue,   M. C. Rodriguez and H. N. Long,
 {\it Nucl. Phys.} {\bf B  870}, (2013)
 293,  [arXiv:1210.6776(hep-ph)] (2012).

\bibitem{s331r} J. C. Montero, V. Pleitez, and M. C. Rodriguez,  Phys. Rev. D {\bf
70}, 075004 (2004).


\bibitem{scalarrhn}  D. T. Huong, M. C. Rodriguez and  H. N. Long, {\it  Scalar sector of   Supersymmetric
 $\mbox{SU}(3)_C\otimes \mbox{SU}(3)_L \otimes \mbox{U}(1)_N$
 Model  with right-handed neutrinos,}  [arXiv: hep-ph/0508045].


\bibitem{marcos}  P. V. Dong, D. T. Huong, M. C. Rodriguez, and H. N. Long,
Eur. Phys. J. C  {\bf 48}, 229 (2006), [arXiv: hep-ph/0604028 ].



\bibitem{susyeco} P. V. Dong, D. T. Huong, M. C. Rodriguez, and H. N.
Long,   Nucl. Phys. {\bf B 772}, 150  (2007); D. T.  Binh, L. T.
Hue, D. T. Huong and  H. N. Long, \emph{ Eur. Phys. J}. \textbf{C 74},  (2014) 2851,
[arXiv:1308.3085(hep-ph)].


\bibitem{higph} P. V. Dong, D. T. Huong, N. T. Thuy, and H. N. Long,
Nucl. Phys.  {\bf B 795}, 361  (2008).

\bibitem{jhep}P. V. Dong, Tr. T. Huong, N. T. Thuy, and H. N. Long,
  JHEP {\bf 11}, 073  (2007).

\bibitem{jhep2} D. T. Huong and H. N. Long,
 JHEP {\bf 07}, 049 (2008), [arXiv:0804.3875(hep-ph)](2008).




\bibitem{infpot} E. J. Copeland, A. R. Liddle, D. H. Lyth, E. D.
Stewart, and D. Wands, Phys. Rev. D {\bf 49}, 6410 (1994).
\bibitem{GDvali} G. Dvali, Q. Shafi and R. Schaefer, Phys. Rev.
Lett. {\bf 73}, 1886 (1994).

\bibitem{linrio} A. Linde and A. Riotto, Phys. Rev. D {\bf 56},
1841 (1997).


\bibitem{Jen} R. Jeannerot, Phys. Rev. D {\bf 56}, 6205 (1997).


\bibitem{Coliman}S. Coleman and S. Weinberg, Phys. Rev. D {\bf 7}, 1888
(1973).


\bibitem{lythrioto} D. H. Lyth and A. Riotto, Phys. Rep. {\bf
314}, 1 (1999).


\bibitem{huong2}  D. T. Huong, H. N. Long, \emph{J. Phys}. G: Nucl. Part. Phys. \textbf{38}, 015202 (2011),
[arXiv:1004.1246(hep-ph)].


\bibitem{hdkt} D. T. Huong, P. V. Dong, C. S. Kim and N. T. Thuy, Phys. Rev. D \textbf{91}, 055023 (2015),
[arXiv:1501.00543(hep-ph)].

\bibitem{majorana} J.  M. Cline, G.  Laporte, H.  Yamashita, S. Kraml, JHEP \textbf{0907}, 040 (2009).
\bibitem{thdm} S. Kanemura, Y. Okada, E. Senaha, Phys. Lett. B \textbf{606}, 361 (2005).
\bibitem{ESMCO} S. W. Ham, S-A Shim, and S. K. Oh, Phys. Rev. D \textbf{81}, 055015 (2010).

\bibitem{elptdm} S.  Das, P.  J. Fox, A. Kumar, and N. Weiner, JHEP  \textbf{1011}, 108 (2010); D. Chung and A. J. Long, Phys. Rev.  D. \textbf{84}, 103513 (2011); M.  Carena, N. R. Shaha, and C. E. M. Wagner, Phys. Rev. D \textbf{85}, 036003 (2012); A.
 Ahriche and S. Nasri, Phys. Rev. D \textbf{85}, 093007 (2012); D. Borah and J. M. Cline, Phys. Rev. D \textbf{86}, 055001 (2013).


\bibitem{SME}K. Kajantie, M. Laine, K. Rummukainen, and M. Shaposhnikov, Phys. Rev. Lett. \textbf{77}, 2887 (1996); F. Csikor, Z. Fodor, and J. Heitger, Phys. Rev. Lett. \textbf{82}, 21 (1999); J. Grant, M. Hindmarsh, Phys. Rev. D. \textbf{64}, 016002 (2001); M. D'Onofrio, K. Rummukainen, A.  Tranberg, JHEP\,
    \textbf{08}, 123 (2012).

\bibitem{BSM}M. Bastero-Gil, C. Hugonie, S. F. King, D. P. Roy, and S. Vempati, Phys. Lett. B \textbf{489}, 359 (2000); A. Menon, D. E. Morrissey, and C. E. M. Wagner, Phys. Rev. D \textbf{70}, 035005 (2004); S. W. Ham, S. K. Oh, C. M. Kim, E. J. Yoo, and D. Son, Phys. Rev. D \textbf{70}, 075001 (2004).



\bibitem{cos2}V. Q. Phong, V. T. Van, and H. N. Long, Phys. Rev. D \textbf{88}, 096009 (2013).

\bibitem{cos4}V. Q. Phong, H. N. Long, V.  T. Van,  L. H.
Minh, \emph{ Eur. Phys. J}.
\textbf{C 75},  (2015) 342,
[arXiv:1409.0750(hep-ph)].

 \bibitem{cos3} V. Q. Phong, H. N. Long, V. T. Van, N. C.
Thanh,  \emph{Phys. Rev.} \textbf{D 90}, 085019 (2014), [arXiv:1408.5657(hep-ph)].
\bibitem{manton} F. R. Klinkhamer and N. S. Manton, Phys. Rev. D \textbf{30} (1984), 2212.

\bibitem{sphasm} T. Akiba, H. Kikuchi, and T. Yanagida, Phys. Rev. D \textbf{40} (1989), 588; F. R. Klinkhamer and N. S.
 Manton, Phys. Rev. D \textbf{30}  (1984), 2212.
\bibitem{spha-nonpur}G.  D. Moore, Phys. Rev. D \textbf{59} (1998), 014503.
\bibitem{Farrar}G.  R. Farrar and M. E. Shaposhnikov, Phys. Rev. Lett. \textbf{70} (1993), 2833.
\bibitem{Arnold} P.  Arnold and L.  McLerran, Phys. Rev. D \textbf{36}  (1987), 581.

\bibitem{ctspha} P. Arnold and L.  McLerran,  Phys. Rev. D \textbf{37} (1988), 1020.
\bibitem{gauge0} Y. Brihaye, J. Kunz, Phys. Rev. D \textbf{48} (1993) 3884.
\bibitem{spha-huble} M. Joyce, Phys. Rev. D \textbf{55}   (1997), 1875.
\bibitem{cross} M. D'Onofrio, K. Rummukainen, A. Tranberg, JHEP {\bf 08} (2012) 123.

\end{thebibliography}
\end{document}